\documentclass[intlimits,twoside,a4paper]{article}

\usepackage[T2A]{fontenc} 
\usepackage[cp1251]{inputenc}
\usepackage[eqsecnum]{cmpj3}

\usepackage{bm}

\issue{2022}{25}{1}{13301}
\doinumber{10.5488/CMP.25.13301}
\title[Investigation of the stability and charge states  of vacancy in clusters Si$_{29}$ and Si$_{38}$ ]%
{Investigation of the stability and charge states of vacancy in clusters Si$_{29}$ and Si$_{38}$}
\author[A.~B. Normurodov \emph{et al.}]{A.~B. Normurodov\orcid{0000-0002-6953-3708}
\thanks{Corresponding author: \email{normurodov@inp.uz, anormurodov@gmail.com}.}, 
       A.~P. Mukhtarov\orcid{0000-0002-1825-0097}, 
       F.~T. Umarova\orcid{0000-0002-1201-0260}, 
       M.~Yu. Tashmetov\orcid{0000-0001-6232-5642},  
       Sh. Makhkamov \orcid{0000-0002-1258-1899}, 
       N.~T. Sulaymonov \orcid{0000-0002-9382-314X}
}
\address{Institute of nuclear physics, 1 Xuroson Str., 100214, Tashkent, Uzbekistan
}

\Keywords{silicon nanoclusters, charge, vacancy, non-conventional tight-binding method, molecular dynamics}

\date{Received May 08, 2021, in final form December 07, 2021}

\begin{document}

\maketitle

\begin{abstract}
Stability and charge states of vacancy in Si$_{29}$ and Si$_{38}$ clusters have been calculated by non-conventional tight-binding method and molecular dynamics. Based on the theoretical calculations, it was shown that the vacancy in pure dimerized clusters is unstable, while in hydrogenated Si$_{29}$H$_{24}$ and Si$_{38}$H$_{30}$ clusters it is stable, but leads to a distortion of its central part with the transition of symmetry from Td to $C_{3v}$ and a change in the forbidden gap. The charges of cluster atoms in the presence of a vacancy are distributed so that all silicon atoms acquire a stable negative charge, which occurs due to the outflow of electrons of the central atom to the neighboring spheres.
\printkeywords
%
%\pacs 36.40.Mr, 61.46.Bc
\end{abstract}

\section{Introduction}

%\doclicenseThis

Silicon nanoparticles as well as crystalline silicon may contain various defects that effect both their optoelectronic and electrophysical properties. The most common defects are the point defects caused by an intrinsic interstitial atom and a vacancy. Moreover, a vacancy can be formed both inside and on the surface of nanoparticles during the crystal growth and under various external influences. The vacancy in crystalline silicon Si was studied in detail. For example, the stable states of the vacancy and its complexes with impurities, their formation energies were theoretically investigated in previous researches~\cite{Luis01,Dabr15,Gyeo02,Yong11,Mani02}. The experimental studies~\cite{Joha00,Kole18,Wat64,Plat21,Emt81} indicated the possibility of forming the complexes of a vacancy with an impurity oxygen and hydrogen atoms in stable states, and the kinetic energies of the formation are given. This type of defects are very mobile and could be identified directly only in the silicon of $p$-type. The vacancy in bulk Si known to form highly localized defect complexes with deep local levels in the band gap of silicon~\cite{Emt81}. The possibility of preserving the energy characteristics of such a defect complex in nanosized particles and the effect of size-dependent phenomena of a particle have not been unambiguously determined.

Controlling the concentration of the vacancy in silicon is one of the key factors in semiconductor electronics. Despite the fact that the vacancy in crystalline silicon has been studied for a long time using various methods, a relatively little attention is paid to the presence and stability of the vacancy in silicon nanoparticles. Although a number of researches on vacancy in nanosilicon have been provided~\cite{Mag18,Xi06,Gui11,Bar19}, detailed studies of its geometric configuration and positions of electronic levels were done only in several of them~\cite{Gui11,Cher16}. Thus, size effects on the vacancy formation energy and entropy were considered in~\cite{Gui11} and it was found that the size reduction makes the vacancy much easier to form; then, the vacancy concentration increases with reducing size and increasing temperature.

Thermodynamic state equation of the nanocrystal with vacancies leads to the conclusion that, regardless of the presence of an external effect on the nanoparticle, even at a constant total temperature of the system of interacting atoms, a certain number of vacancies always formed due to thermal fluctuations~\cite{Cher16}. Therefore, the determination of the structure of the local region of the defect formation with the participation of a vacancy and the effect of the surface on the stability of the defect is of great importance. Zhang et al. researched the quantum confinement effect on the vacancy-induced spin polarization in carbon, silicon, and germanium nanoparticles by density functional analysis~\cite{Zhen10,Khra20}.

This paper presents the results of studies of the structural parameters and energy characteristics of pure and hydrogenated silicon nanoparticles Si$_{29}$ and Si$_{38}$  with vacancies. Such sized clusters are most stable and widely used models of the silicon nanoparticle of one nm size~\cite{Khak07,Oss08}. At present, the computational capabilities make it possible to calculate the main structural and energy parameters of such clusters by ab initio methods. However, there are two very important disadvantages of both groups of Hartree--Fock--Rutaan methods and the methods based on the density functional theory. Taking account of the real conditions for the existence of nanoparticles first of all leads to spending a huge computer time for calculations while a large overestimation of the internal bonds total energy makes it difficult to compare the calculated energy parameters with experimental data. For the above reasons, we decided to apply the non-conventional tight-binding method (NTBM)~\cite{Kha94} to the calculations, which with a correct approach to the parameterization procedure and the choice of the calculation algorithm, makes it possible to achieve a significant reduction in computational costs and a reasonable agreement with experimental data.

\section{Computation method details}

The method for simulation of nanostructures used by us is adapted to optimize the structure and energy
parameters and combines the NTBM proposed by Khakimov~\cite{Kha94,Khak05}, and the molecular
dynamic method.
Here, the expression for the total energy of the NTBM has the form
\begin{align}\label{Eq.1}
	E_{\rm {tot}}=\sum_{\mu}\sum_{\nu>\mu}\frac{Z_{\mu}^{\rm {scr}}Z_{\nu}^{\rm {scr}}}{R_{\mu\nu}}+
	\sum_{\mu}\sum_{\nu>\mu}\frac{Q_{\mu}Q_{\nu}}{R_{\mu\nu}}+
	\sum_{\mu}\sum_{\nu>\mu}\sum_{i}\sum_{j}P_{\mu i,\nu j}H_{\mu
		i,\nu j}+ \sum_{\mu}\left(E_{\mu}-E_{\mu}^{0}\right),
\end{align}
where $R_{\mu\nu}$ is internuclear distance,
\begin{align}\label{Eq.2}
	Z_{\mu}^{\rm {scr}}=Z_{\mu}^{\rm {scr}}\left(R_{\mu\nu,}\{N_{\mu
		i}^{0}\}\right)=Z_{\mu}-\sum_{i}N_{\mu i}^{0} \left[1-\alpha_{\mu i} {\rm{exp}} \left(-\alpha_{\mu
		i}R_{\mu\nu}/R_{\mu}^{0}\right)\right],
\end{align}
\begin{align}\label{Eq.3}
	Q_{\mu}=Z_{\mu}^{\rm {scr}}\left(R_{\mu\nu,}\{N_{\mu
		i}^{0}\}\right)-Z_{\mu}^{\rm {scr}}\left(R_{\mu\nu,}\{N_{\mu
		i}^{0}\}\right)-Z_{\mu}^{\rm {scr}}\left(R_{\mu\nu,}\{N_{\mu i}\}\right)
\end{align}
are screened nuclear and non-point ion charges, respectively; $
Z_{\mu}$ is the charge of the $\mu$-th nucleus or the nucleus with
core electrons; $R_{\mu}^{0}= n/\xi_{\mu i}^{0}$ is the most probable
distance between the nucleus and electron, $n$ and $\xi_{\mu}^{0}$ is the principal quantum number and Slater exponent of the $i$-th atom orbital (AO),
centered on $\mu$-nucleus; $E_{\mu}^{0}$ and $E_{\mu}$ are total
energies of the individual atoms in non-interacting and
interacting systems, characterized by $\{N_{\mu i}^{0}\equiv
P_{\mu i,\mu i,}^{0}\}$ and $\{N_{\mu i}\equiv P_{\mu i,\mu i,}\}$
occupancy numbers and $\{E_{\mu i}^{0}\}$ and $\{E_{\mu i}\}$
corresponding valence AO's energies. AO is supposed to be
orthogonalized, and the matrix equation will be as follows:
\begin{align}\label{Eq.4}
	\sum_{\nu j}(H_{\mu i,\mu j}-\epsilon\delta_{\mu i,\nu j})C_{\nu
		j}=0,
\end{align}
and used to be solved self-consistently for determination of the
${\epsilon_{k}}$ energetic spectra and $C_{\mu j}$ expansion
coefficients of the molecular orbitals (MO) over AO.
Self-consistent calculations are based on iterative recalculation of
the diagonal matrix elements of the Hamiltonian using the
dependency of the bond order matrix
\begin{align}\label{Eq.5}
	P_{\mu i,\nu j}=\sum_{k}N_{k}C_{\mu i}(k)C_{\nu j}(k),
\end{align}
and $N_{\mu i}\equiv P_{(\mu i,\mu i)}$ AO occupancy from $C_{\nu j}$. Here, $k$ denotes a MO and $N_k$ ---
denotes the occupancy number of the $k$-th MO. In conventional tight-binding
method (TBM), the $N_k$ are all equal to $2$ (except the highest MO of a
system with an odd number of electrons). However, NTBM involves a description of charged and excited systems, where one or more MOs may have an occupancy less than 2.

The diagonal and off-diagonal matrix elements of NTBM have the
form
\begin{align}\label{Eq.6}
	P_{\mu i,\mu j}=\left(E_{\mu
		i}-\sum_{\nu\neq\mu}Q_{\nu}/R_{\mu\nu}\right)\delta_{ij},
\end{align}
and
\begin{equation}\label{Eq.7}
\begin{array}{ll}	H_{\mu i,\nu j}=\pm\frac{1}{2}h_{\mu i}h_{\nu
		i}A_{ij}(\vec{R}_{\mu\nu}), & \nu\neq\mu, \end{array} 
\end{equation}
correspondently
\begin{align}\label{Eq.8}
	h_{\mu i}=b_{\mu i}\xi_{\mu i}^{0}{\rm{exp}}\left(-\beta_{\mu
		i}R_{\mu\nu}/\bar{R}_{\mu i}^{0}\right)F_{\mu i},
\end{align}
\begin{align}\label{Eq.9}
	F_{\mu i}=\left\{ 1+{\rm{exp}}\left[-\gamma_{\mu i}\left(R_{\mu\nu}-d_{\mu i}\right)\right]\right\}^{-1},
\end{align}
$\bar{R}_{\mu i}^{0}$ is the average distance between electron and
corresponding nucleus, $A_{ij}(\vec{R}_{\mu\nu})$ are angular
functions tabulated by Slater and Coster~\cite{Khak05}. In~(\ref{Eq.7}), plus sign is taken for $sp$ and $pp-\sigma$ matrix elements, and minus sign is taken for $ss$ and $pp-\pi$ matrix elements.

NTBM yields to traditional TBM in speed due to iterative
self-consistent calculations. To accelerate the convergence of
these calculations, we use the techniques of dynamic damping~\cite{Kha94}
and shift level~\cite{Sla57}. A more rigorous criterion for self-consistency
was used to calculate clusters; the calculations at each point in
configuration-coordinate space are completed only if the values of
AO population in two successive iterations are less than
$10^{9}$.

To determine the possible spatial structures of the system of a
given number of atoms, the molecular dynamics simulation (MD)
method~\cite{All87,Wol99} based on numerical integration of Newton's equations
of motion was used:
\begin{align}\label{Eq.10}
	m_{i}\rd^{2}r_{i}/\rd t^{2}=m_{i}a_{i}=F_{i}; \qquad F_{i}=-\rd U/\rd r_{i},
\end{align}
where $m_{i}$, $r_{i}$ and $a_{i}$ are, accordingly, the mass,
position and acceleration of the $i$-th particle; $F_{i}$ is
force acting to the $i$-th particle by other particles; $U$ is
total potential energy of the system, which can be calculated
using one of the approximate methods.

Equation~(\ref{Eq.10}) can be solved by step using Taylor expansion near
current $t$ time~\cite{Sla57}:
\begin{align}\label{Eq.11}
	r_{t+\delta t}=r_{t}+\nu_{t}\delta t+\frac{1}{2}a_{t}\delta
	t^{2}+\frac{1}{6}b_{t}\delta t^{3}+...,
\end{align}
\begin{align}\label{Eq.12}
	\nu_{t+\delta t}=\nu_{t}+a_{t}\delta t+\frac{1}{2}b_{t}\delta
	t^{2}+\frac{1}{6}c_{t}\delta t^{3}+...,
\end{align}
where $\nu$ is velocity of the particle.

To integrate the equation of motion, a third-order algorithm is
used in which the positions $R$ and particle velocities $v$ are
calculated by the following formulas:
\begin{align}\label{Eq.13}
	R_{t+\delta t}=\left[R_{t}+\nu_{t}\delta
	t+\frac{1}{12}(7a_{t}-a_{t-\delta t})\delta
	t^{2}\right]\cdot \left(\frac{1}{12}\frac{\rd a_{t}}{\rd R_{t}}\delta t^{2}\right)^{-1},
\end{align}
\begin{align}\label{Eq.14}
	\nu_{t+\delta t}=\nu_{t}+\frac{1}{12}(8a_{t}+5a_{t+\delta
		t}-a_{t-\delta t})\delta t,
\end{align}
where $a_{t}$ is acceleration of the particle with $m$ mass at $t$
time moment, $\rd a_{t}/\rd R_{t}$ is a derivative from acceleration over
coordinates of the $R$ particle. In calculating the forces
(accelerations) numerically, their derivatives can be determined
simultaneously using the same two additional calculated values of
the total energy in the positions $R_{t}+\delta R$ and
$R_{t}-\delta R$ ($\delta R$ is small shift):
\begin{align}\label{Eq.15}
	a_{t}=-\frac{1}{m}\frac{E(R_{t}+\delta R)-E(R_{t}-\delta
		R)}{2\delta R},
\end{align}
\begin{align}\label{Eq.16}
	\frac{\rd a_{t}}{\rd t}=-\frac{1}{m}\frac{E(R_{t}+\delta
		R)+E(R_{t}-\delta R)-2E(R_{t})}{\delta R^{t}}.
\end{align}

A self-consistent calculation of the electron density distribution
is made many times for each nuclear configuration of the system,
starting with a trial set of values $\{N_{\mu i}\}$, while the
difference between the input and output values $\{N_{\mu i}\}$ (or
$\{c_{i}\}$) in solving secular equation \ref{Eq.13} gets negligible: \
\begin{align}\label{Eq.17}
	\sum_{\mu i}\sum_{\nu j}(H_{\mu i,\nu
		j}-\epsilon\delta_{ij}\delta_{\mu\nu})c_{\nu j}=0,
\end{align}
where $\{\epsilon_{k}\}$ are electronic energetic levels, $c_{\nu j}$ are expansion coefficients of wave functions for AO, $i$ and $j$ are
atomic orbitals, $H$ is Hamiltonian of the atomic system,
$\delta$ are average quadratic errors.

\section{Results and discussions}

This work presents the results of studying the stability of a
vacancy both in pure Si$_{29}$D and Si$_{38}$D clusters and in
their hydrogenated samples. As a model of a nanoparticle we have
chosen a silicon nanocluster Si$_{29}$D with a dimerized surface
and saturation of the surface $24$ silicon atoms with hydrogen
atoms. This cluster is atomically centered and the symmetry of the
central atom is tetrahedral. While considering the vacancy, the
central atom was removed from the cluster and later the cluster
geometry was optimized.

\begin{figure}[htb]
	\begin{center}
		\includegraphics[width=12cm]{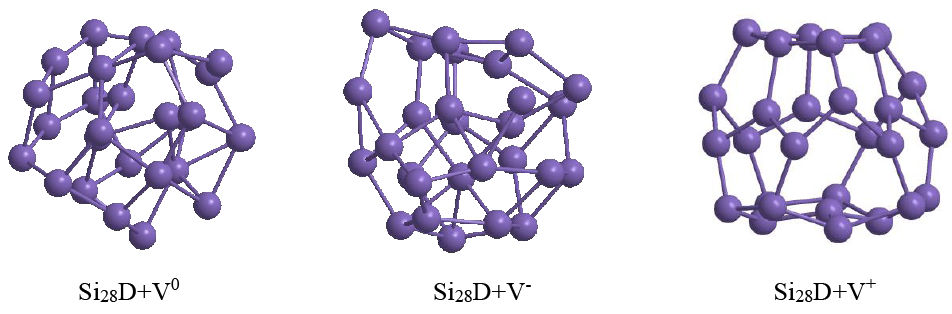}
		\caption{(Colour online) Surface dimerized and optimized cluster structures with vacancy in different charge states.} \label{fig-smp1}
	\end{center}
\end{figure}

The obtained results show (figure~\ref{fig-smp1}) that a vacancy in a pure
Si$_{29}$D cluster is unstable and undergoes a collapse as a result
of a shift of cluster atoms. In this case, the disorder of the
cluster structure increases. Only in the case of a positively
charged  Si$_{29}$D cluster the presence of a vacancy in the
center leads to the formation of a hollow  Si$_{29}$ cluster with
a diameter of 7.04~{\AA}.

\begin{figure}[htb]
	\begin{center}
		\includegraphics[width=5cm]{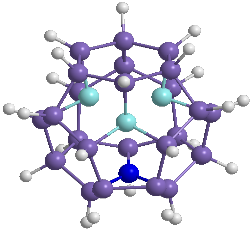}
		\caption{(Colour online) The structure of the vacancy in the dimerized
			Si$_{29}$H$_{24}$ cluster. Silicon atoms are the first neighbors of
			a vacancy and are shown in a color different from the surrounding Si
			atoms (the blue atoms are the atoms shifted to each other, the dark
			blue is a distant atom).}\label{fig-smp2}
	\end{center}
\end{figure}

The obtained dependences of the densities of vacancy energy states
in a dimerized Si$_{29}$D (table~\ref{tbl-smp1}) show that the density of
vacancy states in the neutral dimerized Si$_{29}$D coincides with
the density of states for the pure dimerized Si$_{29}$D without a
vacancy and indicate the metallic nature of conductivity (table~\ref{tbl-smp1}). The HOMO--LUMO gap in these clusters is of the order of $0.01$~eV. In the charged clusters the gap grows.
\begin{table}[htb]
		\caption{The calculated characteristics of the surface dimerized pure silicon cluster with a vacancy in different charge.} \label{tbl-smp1}
	\begin{center}
		\begin{tabular}{|p{2.5cm}|p{2.5cm}|p{2cm}|p{3cm}|p{2cm}}
			\hline
			Clusters & \multicolumn{1}{p{2.2cm}|}{Atomization ener\-gy, eV} &
			\multicolumn{1}{p{3cm}|}{Atomization ener\-gy per atom, eV} &
			\multicolumn{1}{p{2.2cm}|}{Bond length, \AA} &
			\multicolumn{1}{p{2.5cm}|}{HOMO--LUMO gap, eV} \\
			\hline
			Si$_{29}$D & \multicolumn{1}{c|}{124.33} & \multicolumn{1}{c|}{4.28} & \multicolumn{1}{c|}{2.29}& \multicolumn{1}{c|}{0.08}   \\
			Si$_{28}+V^{0}$ & \multicolumn{1}{c|}{120.51} & \multicolumn{1}{c|}{4.30}& \multicolumn{1}{c|}{2.39}& \multicolumn{1}{c|}{0.09}\\
			Si$_{28}+V^{-}$ & \multicolumn{1}{c|}{120.58} & \multicolumn{1}{c|}{4.29} & \multicolumn{1}{c|}{2.38}& \multicolumn{1}{c|}{0.15} \\
			Si$_{28}+V^{+}$ & \multicolumn{1}{c|}{117.75} & \multicolumn{1}{c|}{4.20} & \multicolumn{1}{c|}{2.35}& \multicolumn{1}{c|}{0.26} \\
			\hline
		\end{tabular}
	\end{center}
\end{table}
Investigation of the stability of a neutral vacancy in a
surface-dimerized Si$_{29}$DH$_{24}$ cluster with passivation of
unsaturated bonds has showed that four silicon atoms, the first
neighbors of the vacancy, initially had a symmetrical arrangement
corresponding to the tetrahedral point symmetry group. They
underwent a Jahn--Teller type distortion from the initial positions.
Three atoms as shown in figure~\ref{fig-smp2} approach each other while one of the
Si atoms moves away towards the surface. As a result, the nearest
hydrogen atom connected to the surface atom shifts to a
bonded-centered position in the direction of the remote silicon
atom. The symmetry of the central part of the cluster transfers from the symmetry point group \emph{$T_{d}$} to the symmetry point group \emph{$C_{3v}$} (lowering of the \emph{$T_{d}$} symmetry upon transition to the hexagonal position). In this case, the distance between the three atoms becomes 3.19~{\AA} (ideally 3.75~{\AA}). It
increases 4.12~{\AA} for the removed atom and one of the three connivent atoms. The second neighbors of the vacancy also move away and the distances between the first and second neighbors of the vacancy become 2.28~{\AA}.

\begin{figure}[htb]
	\begin{center}
		\includegraphics[width=15cm]{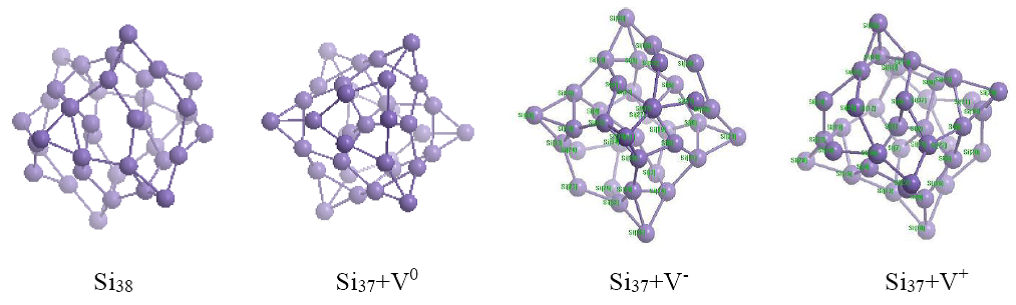}
		\caption{(Colour online) Optimized bare and vacancy-consisted Si$_{38}$  cluster structures.}
	\end{center} \label{fig-smp3}
\end{figure}

The charges of cluster atoms in the presence of a vacancy are
distributed so that all silicon atoms acquire a stable negative
charge that occurs due to the outflow of the electrons of the central
atom to the neighboring spheres. This can be seen from table~\ref{tb1-smp2} where the values of the charges in the two coordination spheres of the
cluster change to a great extent.
\begin{table}[htb]
		\caption{The calculated parameters in the hydrogenated cluster
		Si$_{29}$H$_{24}$ and in the presence of a vacancy.} \label{tb1-smp2}
	\begin{center}
		\begin{tabular}{|p{1.4cm}|p{1.2cm}|p{1.3cm}|p{1.7cm}|p{1.2cm}|p{1cm}|p{1cm}|p{1cm}|p{1.3cm}|}
				\hline
			Clusters & Total ener\-gy, eV  & HOMO--LUMO ener\-gy gap, eV & First Neighbor Shift & \multicolumn{5}{|c|}{Charges} \\
			\cline{5-9}
			& & & & Central atom & $1$ sphere & $2$ sphere & $3$ sphere & Hydrogen atoms\\
			\hline
			$\rm{Si_{29}H_{24}}$ & 181.39 & 1.02 &  & 0.20 & $-0.13$ & $-0.14$ & 0.01 & 0.05 \\
			$\rm{Si_{28}H_{24}V}$ & 173.48 & 0.09 & \emph{$C_{3v}$}
			
			$3x(-0,56)$ $1x(+0,37)$ & $-$ & $-0.08$ & $-0.08$ & $-0.04$ &0.05 \\
				\hline
		\end{tabular}
	\end{center}
\end{table}
In the third sphere, the charge changes are insignificant
compared to the cluster without a vacancy. The presence of a
vacancy leads to a substantial rearrangement of the bonds among
the nearest neighbors of the vacancy and gives a number of energy
levels in the gap between the HOMO and LUMO. Apparently, this is
due to the approach of the three atoms and the formation of a weak
covalent bonds between them. At the same time, the fourth atom
tends to float to the surface at an equal distance from the
center of the cluster. To determine the effect of a vacancy for
the structural rearrangement of the Si$_{38}$ cluster, the
equilibrium geometries and the energy parameters of a number of
clusters were calculated. We have investigated those for dimerized
Si$_{38}$ cluster without a vacancy, three charge states $(0, +,
-)$ of the Si$_{37}$ cluster with a vacancy in the center formed
from the Si$_{38}$ cluster by the removal of one of the central atoms.
MD optimized structures of the pure~Si$_{38}$ cluster and with a
vacancy in the center in different charge states are shown in
figure~\ref{fig-smp3}. Table~\ref{tb1-smp3} shows the atomization energies and the HOMO--LUMO gap.
As can be seen from figure~\ref{fig-smp3}, the cluster structure during the
formation of a vacancy retains a tetrahedral symmetry of the cubic system. However,
the volume of the central part of clusters with a vacancy
decreases because of an increase in the length of Si-Si bonds
between atoms. The number of possible configurations of these
clusters containing a vacancy in the center is equal to the number
of hexagonal positions around the tetrahedral center, i.e., equal
to 6. With the removal of one silicon atom from the Si$_{38}$
cluster, the atomization energy slightly decreases, and the most
stable is a negatively charged cluster with a vacancy in the
center. The HOMO--LUMO gap increases by a factor of 2 for
Si$_{37}+V^{0}$, and decreases by a factor of 2 for
Si$_{37}+V^{+}$. The analysis of the components of the atomization
energy (table~\ref{tbl-smp4}) shows that, in the presence of a vacancy, the
contribution of the ion-ion interaction sharply increases for a
positively charged state. 

\begin{table}[h]
	\caption{The calculated characteristics of vacancy-consisted Si$_{38}$D cluster in different charge states.} \label{tb1-smp3}
	\begin{center}
		\begin{tabular}{|p{2.5cm}|p{2.5cm}|p{2cm}|p{3cm}}
			\hline
			Clusters & \multicolumn{1}{p{1.8cm}|}{Si-Si~bond length, \AA}
			& \multicolumn{1}{p{3cm}|}{Atomization ener\-gy per atom, eV} &
			\multicolumn{1}{p{2.5cm}|}{HOMO--LUMO gap, eV} \\
			\hline
			Si$_{38}$D & \multicolumn{1}{c|}{2.28} & \multicolumn{1}{c|}{4.25} & \multicolumn{1}{c|}{0.15}   \\
			Si$_{37}+V^{0}$ & \multicolumn{1}{c|}{2.35} & \multicolumn{1}{c|}{4.16}& \multicolumn{1}{c|}{0.30}\\
			Si$_{37}+V^{-}$ & \multicolumn{1}{c|}{2.33} & \multicolumn{1}{c|}{4.17} & \multicolumn{1}{c|}{0.22} \\
			Si$_{37}+V^{+}$ & \multicolumn{1}{c|}{2.36} & \multicolumn{1}{c|}{4.14} & \multicolumn{1}{c|}{0.08} \\
			\hline
		\end{tabular}
	\end{center}
\end{table}

Accordingly, the bond between the atoms
of the cluster core and surface atoms is weakened. Since in all
cases the positive charge of the cluster surface is preserved, it
can be assumed that this occurs due to a change in the electron
densities inside the clusters.

\begin{table}[htb]

	\caption{Components of atomization ener\-gy of clusters consisting of 38 atoms with a vacancy in different charge states.} \label{tbl-smp4}
	\begin{center}
		\begin{tabular}{|p{2cm}|p{2.5cm}|p{2cm}|p{3cm}|p{3cm}}
				\hline
			Dimerized cluster & \multicolumn{1}{p{2.5cm}|}{Atomization
				ener\-gy, eV} & \multicolumn{1}{p{2.5cm}|}{Atom-atom inte\-raction,
				eV} & \multicolumn{1}{p{3cm}|}{Binding ener\-gy, eV} &
			\multicolumn{1}{p{2cm}|}{Ion-ion inte\-raction, eV} \\
			\hline
			Si$_{38}$ & \multicolumn{1}{c|}{161.39} & \multicolumn{1}{c|}{48.24} & \multicolumn{1}{c|}{286.54}& \multicolumn{1}{c|}{0.90}   \\
			Si$_{38}^{-}$ & \multicolumn{1}{c|}{165.31} & \multicolumn{1}{c|}{48.54}& \multicolumn{1}{c|}{287.19}& \multicolumn{1}{c|}{0.91}\\
			Si$_{38}^{+}$ & \multicolumn{1}{c|}{154.77} & \multicolumn{1}{c|}{47.99} & \multicolumn{1}{c|}{286.43}& \multicolumn{1}{c|}{2.99} \\
			Si$_{37}+V^{0}$ & \multicolumn{1}{c|}{154.13} & \multicolumn{1}{c|}{47.21} & \multicolumn{1}{c|}{280.56}& \multicolumn{1}{c|}{0.77} \\
			Si$_{37}+V^{-}$ & \multicolumn{1}{c|}{154.22} & \multicolumn{1}{c|}{46.08}& \multicolumn{1}{c|}{277.15}& \multicolumn{1}{c|}{0.31}\\
			Si$_{37}+V^{+}$ & \multicolumn{1}{c|}{153.22} & \multicolumn{1}{c|}{44.03} & \multicolumn{1}{c|}{273.06}& \multicolumn{1}{c|}{2.90} \\
				\hline
		\end{tabular}
	\end{center}
\end{table}

\begin{figure}[htb]
		\begin{center}
		\includegraphics[width=13cm]{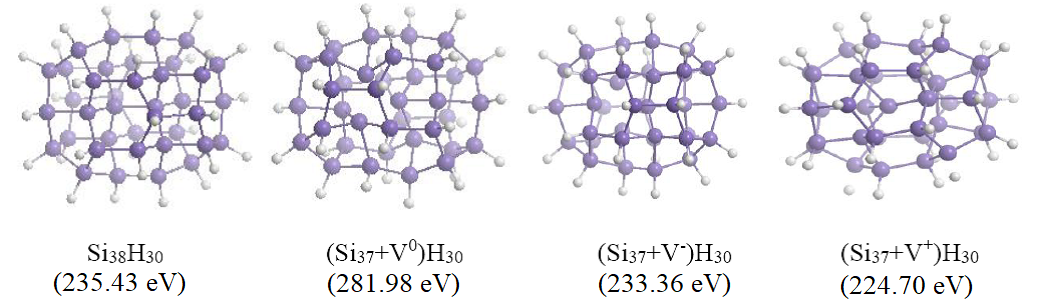}
		\caption{(Colour online) Geometry of optimized Si$_{38}$ clusters in different
			charge states with complete passivation of surface bonds.} \label{fig-smp4}
	\end{center} 
\end{figure}

The study of hydrogen passivated clusters in the presence of a
vacancy shows that complete hydrogen passivation of dangling bonds
on the surface stabilizes (freezes) the structure of clusters with the
surface reconstruction. Figure~\ref{fig-smp4} comparatively shows the structure
and atomization energies of the studied clusters. As can be seen
from figure~\ref{fig-smp4}, the presence of a vacancy distorts the cluster
structure while maintaining the tetrahedral symmetry of the bonds.
A neutral, passivated by hydrogen, Si$_{38}$ cluster containing a
vacancy is more stable than a cluster without a vacancy, the
difference being 46.55~eV. In the case of the (Si$_{37}+V^{+}$)H$_{30}$ cation, the cluster volume increased by about 10\%, and
the central atom shifted toward the vacancy with the formation of
a bond with the peripheral silicon atom.
The values of atomization energies are given in parentheses. Table~\ref{tbl-smp5} shows the atomization energy and the HOMO--LUMO gap for passivated clusters. As can be seen, in general, the passivation of hydrogen by clusters containing a vacancy does not change the
atomization energy with the exception of (Si$_{37}+V^{0}$)H$_{30}$, but the gap significantly decreases compared to a cluster without a vacancy.
\begin{table}[htb]
	
	\caption{Atomization energies and gap widths of hydrogen-passivated and vacancy-consisted clusters in different charge states.} 	\label{tbl-smp5}
	\begin{center}
		\begin{tabular}{|p{3cm}|p{4cm}|p{4cm}}
				\hline
			Clusters & \multicolumn{1}{p{3cm}|}{Atomization ener\-gy per atom,
				eV} & \multicolumn{1}{p{3cm}|}{HOMO--LUMO gap, eV} \\
			\hline
			Si$_{38}$H$_{30}$ & \multicolumn{1}{c|}{3.46} & \multicolumn{1}{c|}{0.92}   \\
			(Si$_{37}+V^{0}$)H$_{30}$ & \multicolumn{1}{c|}{4.21} & \multicolumn{1}{c|}{0.18} \\
			(Si$_{37}+V^{-}$)H$_{30}$ & \multicolumn{1}{c|}{3.46} & \multicolumn{1}{c|}{0.24} \\
			(Si$_{37}+V^{+}$)H$_{30}$ & \multicolumn{1}{c|}{3.47} & \multicolumn{1}{c|}{0.17}\\
				\hline
		\end{tabular}
	\end{center}
\end{table}
Among the clusters containing a vacancy, the largest HOMO--LUMO gap is observed in a negatively charged cluster. Thus, in the presence of a vacancy, the cluster structure is distorted, and the electronic properties change.

\section*{Conclusions}

Thus, based on the theoretical calculations, it was shown that the vacancy in pure dimerized clusters is unstable, while in hydrogenated Si$_{29}$H$_{24}$ and Si$_{38}$H$_{30}$ clusters it is stable, but it leads to a distortion of its central part with the transition of symmetry from $T_{d}$ to $C_{3v}$ and a change in the forbidden gap. The presence of a vacancy leads to a substantial rearrangement of the bonds between the nearest neighbors of the vacancy and gives a number of energy levels to the gap region between the HOMO and LUMO. Apparently, this is due to the approach of the three atoms and the formation of a weak covalent bond between them, while the fourth atom appears to float to the surface at the same distance from the center of the cluster. The charges of cluster atoms in the presence of a vacancy are distributed so that all silicon atoms acquire a stable negative charge, which occurs due to the outflow of electrons of the central atom to the neighboring spheres.

%
%% If you have problems with typesetting in ukrainian uncomment lines below.
%
%  \lastpage
%  \end{document}
\newpage

\ukrainianpart

\title{Дослідження стійкості і зарядових станів вакансії в кластерах Si$_{29}$ і Si$_{38}$}

\author{А.~Б. Нормуродов\orcid{0000-0002-6953-3708},
	А.~П. Мухтаров\orcid{0000-0002-1825-0097},
	Ф.~Т. Умарова\orcid{0000-0002-1201-0260},
	М.~Ю. Ташметов\orcid{0000-0001-6232-5642}, 	Ш. Махкамов \orcid{0000-0002-1258-1899},
	Н.~Т. Сулеймонов \orcid{0000-0002-9382-314X}
}
\address{Інститут ядерної фізики, вул. Хуросонська 1, 100214, Ташкент, Узбекистан}
\makeukrtitle

\begin{abstract}
	За допомогою нестандартного методу сильного зв'язку та молекулярної динаміки розраховані стійкість і зарядові стани вакансії в кластерах Si$_{29}$ і Si$_{38}$. На основі теоретичних розрахунків було показано, що вакансія в чистих димеризованих кластерах є нестійкою, в той час як в гідрогенізованих  Si$_{29}$H$_{24}$ та Si$_{38}$H$_{30}$ кластерах вона є стійкою, але призводить до деформації центральної частини кластерів зі зміною симетрії з Td до $C_{3v}$ та до змін в забороненій зоні. Заряди атомів кластера при наявності вакансії розподілені таким чином, що всі атоми кремнію набувають стійкого від'ємного заряду, що відбувається завдяки відтоку електронів центрального атома на сусідні сфери.
	\keywords нанокластери кремнію, заряд, вакансія, нестандартний метод сильного зв'язку, молекулярна динаміка
\end{abstract}

\lastpage
\end{document}